\documentclass[12pt]{article}
\usepackage{graphicx}
\thicklines

\def\be{\begin{equation}}
\def\ee{\end{equation}}
\newcommand{\bea}{\begin{eqnarray}}
\newcommand{\eea}{\end{eqnarray}}

\def\3i{\int\!\!\!\int\!\!\!\int}
\def\2i{\int\!\!\!\int}

\def\ddel{{}^\bullet\! \Delta}
\def\deld{\Delta^{\hskip -.5mm \bullet}}
\def\dddel{{}^{\bullet \bullet} \! \Delta}
\def\ddeld{{}^{\bullet}\! \Delta^{\hskip -.5mm \bullet}}

\def\la{\langle}
\def\ra{\rangle}

\def\t{\tau}

\def\rldd{\rlap{\,/}\nabla}

\usepackage{graphicx}
\usepackage{amsfonts}

\newcommand{\EQ}{\begin{equation}}
\newcommand{\EN}{\end{equation}}

\def\a{\alpha}

\def\w{\omega}
\def\p{\partial}

\def\sqr#1#2{{\vcenter{\vbox{\hrule height.#2pt
     \hbox{\vrule width.#2pt height#1pt \kern#1pt
           \vrule width.#2pt}
       \hrule height.#2pt}}}}

\begin{document}

\begin{flushright}
\begin{minipage}{0.25\textwidth} hep-th/0211134
\end{minipage}
\end{flushright}
\begin{center}
\bigskip\bigskip\bigskip
\baselineskip=24pt
{\bf\Large{Dimensional regularization for ${\cal N}=1$ susy sigma models 
and the worldline formalism}}
\baselineskip=18pt
\vskip 1cm
\bigskip
Fiorenzo Bastianelli \footnote{E-mail: bastianelli@bo.infn.it}, 
Olindo Corradini \footnote{E-mail: corradini@bo.infn.it} 
and 
Andrea Zirotti \footnote{E-mail: zirotti@bo.infn.it} 
\\[.4cm]
{\em Dipartimento  di Fisica, Universit\`a di Bologna 
and  INFN, Sezione di Bologna\\ 
via Irnerio 46, I-40126 Bologna, Italy}
\end{center}
\baselineskip=18pt
\vskip 2.3cm

\centerline{\large{\bf Abstract}}
\vspace{.4cm}

We generalize the worldline formalism to include spin 1/2 fields 
coupled to 
gravity. To this purpose we first extend dimensional regularization to 
supersymmetric nonlinear sigma models in one dimension. 
We consider a finite propagation time and find that dimensional regularization
is a manifestly supersymmetric regularization scheme, since the  classically 
supersymmetric action does not need any counterterm to preserve worldline 
supersymmetry. We apply this regularization scheme to the worldline 
description of Dirac fermions coupled to gravity. 
We first compute the trace anomaly of a Dirac fermion in 4 dimensions,
providing an additional check on the regularization 
with finite propagation time. Then we come to the main topic and consider 
the one-loop effective action for a Dirac field in a gravitational 
background. We describe how to represent this effective action as a worldline 
path integral and compute explicitly the one- and two-point 
correlation functions, i.e. the spin 1/2 particle contribution to the 
graviton tadpole and graviton self-energy.
These results are presented for the general case of a massive fermion.
It is interesting to note that in the worldline formalism
the coupling to gravity can be described entirely in terms
of the metric, avoiding the introduction of a vielbein.
Consequently, the fermion--graviton vertices are always 
linear in the graviton, just like the standard coupling of fermions 
to gauge fields. 
\newpage

\section{Introduction}

One dimensional supersymmetric nonlinear sigma models are useful to describe 
in first quantization the propagation of fermionic particles in a curved 
background. In fact, it is well-known that ${\cal N}=1$ supersymmetric 
sigma models describe the worldline dynamics of a spinning particle 
 \cite{Brink:1976sz}.
Mastering the path integral quantization of such models provides
a useful tool for treating spin 1/2 particles coupled to gravity.
The purpose of this paper is twofold. 
We first extend dimensional regularization to supersymmetric sigma models 
with finite propagation (proper) time. 
Then, with this regularization scheme at hand,
we generalize the worldline formalism to include spin 1/2 fields
coupled to gravity. This extends the scalar particle case treated in 
\cite{Bastianelli:2002fv}.
The resulting Feynman rules are simpler than the standard ones obtained
from the second quantized action. 
In particular, the fermion--graviton vertices 
can always be taken linear in the graviton field, a fact which seems 
to point once more to unexpected perturbative relations between gravity 
and gauge theories, as reviewed in \cite{Bern:2002kj}.

Path integrals for supersymmetric sigma models in one dimensions were 
originally used for deriving formulas for index theorems and chiral anomalies
\cite{Alvarez-Gaume:1983at,Alvarez-Gaume:1983ig,Windey:1983se}.
However, for obtaining those results the details of how to properly define 
and regulate the path integrals at higher loops were not necessary. 
Due to the worldline supersymmetry the chiral anomalies are seen as
a topological quantity, the Witten index
\cite{Witten:df}, which is independent
of $\beta$, the propagation time in the sigma model.
Thus a semiclassical approximation (which consists in calculating 
a few determinants) already gives the complete results.
The quantum mechanical calculation of chiral anomalies can be extended
to trace anomalies \cite{Bastianelli:1992be,Bastianelli:1993ct}.
However, in the latter case the details of how to define the path integral
is essential since one-loop (in target space) trace anomalies
correspond to higher-loop calculations on the worldline, namely
the one-loop trace anomaly in $D$ dimensions is given by a ${D\over 2}+1$ 
loop calculation on the worldline.
Several regularization schemes have been developed for this purpose:
mode regularization (MR) 
\cite{Bastianelli:1992be,Bastianelli:1993ct,Bastianelli:1998jm}, 
time slicing (TS) \cite{deBoer:1995hv,deBoer:1995cb}, and 
dimensional regularization (DR) \cite{Bastianelli:2000nm,Bastianelli:2000dw}.
The DR regularization was developed after the results 
of \cite{KC} which dealt with nonlinear sigma model 
in the infinite propagation time 
limit \footnote{Recently, Kleinert and Chervyakov \cite{Kleinert:2002zp}
have also analyzed nonlinear sigma models for finite propagation time, 
discussing how DR defines products of distributions, and finding results 
for the Feynman rules which agree with those obtained 
in \cite{Bastianelli:2000nm}.}.
The first objective of this paper is to
extend dimensional regularization 
to include fermionic fields on the worldline and treat 
supersymmetric nonlinear sigma models.
Worldline fermions coupled to gravity give rise to new (superficially)
divergent Feynman diagrams, other than those associated to the coupling 
of gravity with the bosonic coordinates. Hence, one may {\em a priori} expect
additional counterterms to arise.
In fact, in time slicing, the inclusion of the fermionic fields
brings in additional non-covariant counterterms of order $\beta^2$: \
they are proportional to 
$g^{\mu\nu}\Gamma^\lambda_{\mu \rho}\Gamma^\rho_{\nu \lambda}$ 
if one uses fermions with curved target space indices,
or $ g^{\mu\nu} \omega_\mu{}^{ab} \omega_{\nu ab}$
if one uses fermions with flat space indices \cite{deBoer:1995cb}.
Note that such counterterms only arise at two loops, and thus they do not affect
the calculation of the chiral anomalies, but should be included 
if one wants to check with TS that there are no higher 
order corrections in $\beta$ \cite{Waldron:1995tq}.
We are going to show that in dimensional regularization no
extra counterterms arise. This implies that dimensional regularization 
manifestly preserves supersymmetry. In fact the bosonic part produces a
coupling to the scalar curvature with the precise coefficient
required by supersymmetry. We describe how to use both flat 
target space indices and curved ones, for the fermionic fields.
Using curved indices will bring in a new set of bosonic
``ghost'' fields, in the same fashion of   
\cite{Bastianelli:1992be,Bastianelli:1993ct}.

Having at hand a simple and reliable regularization scheme
for supersymmetric sigma models, we turn to the worldline formalism.
As a warm up, we first compute the trace anomaly for a Dirac fermion 
in 4 dimensions. We obtain the expected result, 
providing a further test on our application of the DR scheme.
Then we come to the core of our paper: 
the generalization of the worldline formalism 
to include spin 1/2 particles coupled to gravity.
Many simplifications are known to occur in the  
worldline path integral formulation of quantum field theory,
which for this very reason provides an efficient and alternative method
for computing Feynman diagrams.
This method has quite a long history, rooted  in \cite{Feynman50}.
Later it was developed further by viewing it as 
the particle limit of string theory \cite{BK}, and then discussed directly as 
the first quantization of point particles \cite{Polyakov,Strassler}
(see \cite{Schubert:2001he} for a review and a list of references).
The inclusion of background gravity was
presented in \cite{Bastianelli:2002fv} for the case of a scalar particle.
Results  obtained using string inspired rules with gravity
were presented in \cite{Bern:1993wt,Bern:1998ug,Bern:2002kj}.

Here we consider the case of the one-loop effective action for a 
Dirac fermion in a gravitational background.
We describe how to represent it as a worldline path integral. 
We compute explicitly the one- and two-point 
correlation functions, i.e. the spin 1/2 particle contribution to the 
graviton tadpole and graviton self-energy.
These results are presented for the general case of a massive fermion.
In our calculations we use the DR scheme 
constructed in the previous sections. 
The other known scheme explicitly developed to
include worldline fermions (time slicing \cite{deBoer:1995cb}) 
can be used as well, but lack of manifest covariance makes its use 
more complicated.
It is interesting to note that in the worldline formalism
the coupling to gravity can be described entirely in terms
of the metric, avoiding the introduction of a vielbein.
The fermion--graviton vertices are always 
linear in the metric field, just like the standard coupling of fermions 
to gauge fields are linear in the gauge potential. 
This fact seems to point once more to the unexpected perturbative 
relations between gravity and gauge theories encoded in the 
so-called KLT relations \cite{Kawai:1985xq}, as reviewed in \cite{Bern:2002kj}.

The paper is organized as follows. In section 2 we introduce
dimensional regularization applied to the worldline Majorana fermions
and to supersymmetric sigma models. 
We mainly consider antiperiodic boundary conditions
(which break supersymmetry), but also briefly discuss 
periodic boundary conditions.
In section 3 we apply DR to compute the trace anomaly of a Dirac field 
in 4 dimension with quantum mechanics.
Then in section 4 we describe the worldline formalism 
with Dirac fields coupled to gravity and compute explicitly
the spin 1/2 particle contribution to the graviton tadpole and graviton 
self-energy. 
This is the main section, and the reader uninterested in the details 
of the DR regularization scheme may jump directly to it.
Section 5 contains our conclusions. Conventions and useful formulas are 
collected in the appendix. 
We work with an euclidean time both on the worldline and in target space.
The latter is assumed to have even dimensions $D$.

\section{Dimensional regularization with fermions}

In this section we describe the dimensional regularization of fermionic 
path integrals obtained by extending the method presented
in \cite{Bastianelli:2000nm} for bosonic models.
We shall discuss explicitly path integrals for Majorana fermions
on a circle with antiperiodic boundary conditions (ABC),
as these are the only boundary conditions that will be directly 
needed in the applications to trace anomalies and effective action 
calculations.
Our strategy will be as follows: we first set up the rules
of dimensional regularization for fermions following 
\cite{Bastianelli:2000nm}, then we require that a two-loop 
computation with DR reproduces known results, and precisely 
those obtained by a path integral with time slicing \cite{deBoer:1995cb}
(or equivalently by heat kernel methods \cite{DW}).
This requirement plays the role of a standard (in QFT) renormalization
condition, and fixes once for all the DR two-loop counterterm due to fermions.
Since counterterms are due to ultraviolet effects, 
the infrared vacuum structure and the related boundary conditions on the 
fields should not matter in their evaluation. 
Therefore one expects that the same counterterm should apply
to fermionic path integral with periodic boundary conditions (PBC) 
as well. 
No higher-loop contributions to the counterterm are expected
as the model is super-renormalizable, just as in the purely bosonic case. 

Let us consider the path integral quantization of the 
${\cal N}=1$ supersymmetric model
written in terms of Majorana fermions with flat target space indices
\bea 
Z \!\!&=&\!\! 
\int\! Dx Da Db Dc D\psi \ e^{-S} \\
S \!\!&=&\!\!  {1\over \beta}
\int_{0}^1 \!\!
d\tau\, \Big [
{1\over 2}g_{\mu\nu}(x) (\dot x^\mu \dot x^\nu + a^\mu a^\nu  + b^\mu c^\nu )
+ {1\over 2}  \psi_a (\dot \psi^a + \dot x^\mu 
\omega_\mu{}^a{}_b(x) \psi^b)
\nonumber \\
&+& \beta^2 (V(x)+V_{CT}(x) + V_{CT}'(x))
\Big ] 
\label{4.4.1}
\eea
where as usual we have scaled the propagation time $\beta$ out of the action.
The propagation time $\beta$ 
will be considered as the expansion parameter for a perturbative evaluation
(i.e. the loop counting parameter).
In the action we have included:
$i)$ 
the bosonic $a^\mu$ and fermionic $b^\mu,c^\mu$ ghost fields which 
exponentiate the nontrivial path integral measure,
$ii)$ 
the counterterm $V_{CT}$ which arises in the
chosen regularization scheme from
the bosonic sector and which is fixed in order 
to produce a quantum hamiltonian without nonminimal coupling
to the scalar curvature $R$, 
$iii)$ 
the additional counterterm $V_{CT}'$ which may arise 
from the fermionic sector, and
$iv)$
the potential $V= {1\over 8} R$ which is required to have a supersymmetric 
quantum hamiltonian as given by the square of the supersymmetry charge.

The action is classically supersymmetric if all the potential terms
multiplied by $\beta^2$ are set to zero (the ghosts can be trivially
eliminated by using their algebraic equations of motion). 
Supersymmetry may be broken by boundary conditions,
e.g. periodic for the bosons and antiperiodic for the fermions.
Here we assume antiperiodic boundary conditions (ABC) 
for the Majorana fermions $\psi^a(1) =- \psi^a(0)$.
Majorana fermions realize the Dirac gamma matrices in a path integral context,
and ABC compute the trace over the Dirac 
matrices. For simplicity we consider a target space with even dimensions $D$,
and thus the curved indices $\mu,\nu,...$ and the flat space indices
$a,b,...$ both run from 1 to $D$.

One may explicitly compute by time slicing the transition amplitude
for going from the background point $x_0$ at time $t=0$ 
back to the same point $x_0$ at a later time $t=\beta$ using ABC for the 
Majorana fermions. In the two-loop approximation this calculation gives
\bea
Z\equiv{\rm tr}\, \la x_0| e^{-\beta \hat H}|x_0\ra 
= {2^{D\over 2}
\over (2 \pi \beta)^{D\over 2}} 
\Big (1 - {\beta\over 24} R + O(\beta^2)\Big )
\label{4.4.2}
\eea
where the trace on the left-hand side is only over the Dirac matrices, 
and where 
\bea
\hat H = -{1\over  2}\rldd \rldd 
= -{1\over  2}  \nabla^2 + {1\over 8} R
\label{4.4.ham}
\eea
is the supersymmetric Hamiltonian  of the ${\cal N}=1$ model
(one can normalize the supersymmetric charge as 
$\hat Q = {i\over \sqrt{2}} \rldd $, so that
 $ \hat H = \hat Q^2$).
Note that there is an explicit coupling to the scalar curvature
in (\ref{4.4.ham}), thus one needs to use a potential  $V= {1\over 8} R$
in the action together with the time slicing counterterms
$V_{TS} =-{1\over 8}R +{1\over 8}g^{\mu\nu}\Gamma^\lambda_{\mu \rho}
\Gamma^\rho_{\nu \lambda}$ 
and $V_{TS}'= {1 \over 16} g^{\mu\nu} \omega_\mu{}^{ab} \omega_{\nu ab}$
(see \cite{deBoer:1995cb}; later on we will derive once more
this value of $V_{TS}'$ as well).
Our conventions for the curvature tensors can be found in 
section A.1 of the appendix.

Now we want to reproduce eq. (\ref{4.4.2}) 
in dimensional regularization with a
path integral over Majorana fermions.
This will unambiguously fix the additional counterterm $V_{DR}'$
due to the fermions.
Note that in dimensional regularization the potential $V={1\over 8} R$  
cancels exactly with the counterterm  $V_{DR}=-{1\over 8} R$ 
coming from the bosons \cite{Bastianelli:2000nm}.

We focus directly on the regularization of the
Feynman graphs arising in perturbation theory.
To recognize how to dimensionally continue the various Feynman graphs
we extend the action in (\ref{4.4.1}) from 1 to $d+1$ dimensions as follows
\bea
 S = {1\over \beta}
\int_\Omega \!\! d^{d+1}t \, \Big [{1\over 2}\, g_{\mu\nu}
(\partial^\alpha x^\mu \partial_\alpha x^\nu +a^\mu a^\nu +b^\mu c^\nu)
+ {1\over 2}  \bar \psi_a 
\gamma^\alpha (\partial_\alpha 
\psi^a + \partial_\alpha  x^\mu \omega_\mu{}^a{}_b \psi^b ) 
+ \beta^2  V_{DR}' \Big ] 
\label{4.4.3}
\eea
where 
$ \Omega =I \times R^d$ is the region of
integration containing the finite interval $I=[0,1]$,
$\gamma^\alpha$ are the  gamma matrices in $d+1$ dimensions
 satisfying
 $\{ \gamma^\alpha, \gamma^\beta\}= 2\delta^{\alpha\beta}$, and
  $t^\alpha \equiv (\t, {\bf t})$ with $\alpha=0,1,\ldots,d$ and with
a bold face indicating vectors in the extra $d$ dimensions.
Here we assume that we can first continue to those Euclidean integer dimensions
where Majorana fermions can be defined. The Majorana conjugate
is defined by $\bar \psi_a = \psi_a^T C_\pm $ 
with a  suitable charge conjugation matrix $C_\pm $ 
such that $\bar \psi^a \gamma^\alpha \psi^b= -\bar \psi^b \gamma^\alpha 
\psi^a$. 
This can be achieved for example in 2 
dimensions \footnote{In Euclidean 2 dimensions
one can choose $\gamma^1=\sigma^3$, $\gamma^2=\sigma^1$ and $C_+= 1$. 
Recall that $C_\pm$ are defined by
$C_\pm \gamma^\mu C_\pm^{-1} = \pm  
\gamma^{\mu\, T}$.}.
It realizes the basic requirement for the Majorana 
fermions of the ${\cal N}=1$ supersymmetric model 
which must have a non-vanishing coupling
$\omega_{\mu ab}\psi^a\psi^b=-\omega_{\mu ab}\psi^b\psi^a$.
The actual details of how to represent $C_\pm$ and the gamma matrices
in $d+1$ dimensions are not important, as the most important thing for the
rules which define the DR scheme for fermions 
is to keep track how derivatives are going to be contracted 
in higher dimensions.
Apart from the above requirements, no additional Dirac algebra on the gamma
matrices $\gamma^\alpha$ in $d+1$ dimensions is needed.
With these rules one can recognize from the action
(\ref{4.4.3}) the propagators and vertices in $d+1$ dimensions, and thus
rewrite those Feynman diagrams which are ambiguous in one dimension 
directly in $d+1$ dimensions.

The bosonic and ghost propagators are as usual and reported 
in section A.2 of the appendix.
The fermionic fields with ABC on the worldline, $\psi^a(1) =- \psi^a(0)$,
can be expanded in half-integer modes
\be
\psi^a
(\tau) = \sum_{r\in Z+{1\over 2}}
\psi^a_r \, {\rm e}^{2  \pi i r \tau} 
\label{4.4.4}
\ee
and have the following unregulated propagator
\be
\la \psi^a(\tau) \psi^b(\sigma)\ra 
= \beta \delta^{ab} \Delta_{AF}(\tau-\sigma)\ ,
\ \ \ \ 
\Delta_{AF}(\tau-\sigma) =
\sum_{r\in Z+{1\over 2}} 
{1\over 2 \pi i r}
 {\rm e}^{2  \pi i r (\tau-\sigma)} \ .
\label{4.4.5}
\ee

Note that the Fourier sum defining the function
$\Delta_{AF} $ for the antiperiodic fermions
is conditionally convergent for $\tau\neq \sigma$,
and yields
\be
\Delta_{AF}(\tau-\sigma) 
= {1\over 2}
\epsilon (\tau-\sigma) 
\label{4.4.6}
\ee
where $\epsilon (x) =\theta(x) - \theta(-x)$ 
is the sign function (with the value $\epsilon (0)=0$, obtained
by symmetrically summing the Fourier series).
The function $\Delta_{AF} $ 
satisfies 
\be
\partial_\tau \Delta_{AF}(\tau-\sigma) =
\delta_{A}(\tau-\sigma)
\label{4.4.7}
\ee
where $\delta_{A}(\tau-\sigma)$ is the Dirac's delta on functions with 
antiperiodic boundary conditions
\be
\delta_{A}(\tau-\sigma)=
\sum_{r\in Z+{1\over 2}}  {\rm e}^{2 \pi i r (\tau-\sigma)} \ . 
\label{4.4.8}
\ee

The dimensionally regulated propagator obtained by adding a number $d$ of
extra infinite coordinates is derived from (\ref{4.4.3}) and reads  
\be
\la \psi^a(t) \bar \psi^b(s)\ra =
\beta\, \delta^{ab} \Delta_{AF}(t,s) 
\label{4.4.9}
\ee
where the function
\bea 
\Delta_{AF}(t,s)
= - i \int {d^d{\bf k}\over (2\pi)^d} 
\sum_{r\in Z+{1\over 2}} 
 { 2 \pi r \gamma^0 
+ {\bf k \cdot \vec \gamma } \over (2 \pi r)^2+{\bf k}^2}\,
 {\rm e}^{2  \pi i r (\tau-\sigma)}
{\rm e}^{i{\bf k}\cdot ({\bf t}-{\bf s})}  
\label{4.4.10}
\eea
satisfies
\be
\gamma^\alpha {\partial\over \partial t^\alpha} \Delta_{AF}(t,s)=
-{\partial\over \partial s^\beta} \Delta_{AF}(t,s) \gamma^\beta 
=\delta_{A}(\tau-\sigma) \delta^d({\bf t-s}) \equiv
\delta_A^{d+1}(t-s) 
\ .
\label{4.4.11}
\ee
The latter are the basic relations which will be used in the application
of DR to fermions. They keep track of which derivative
can be contracted to which vertex to produce the $d+1$ delta function.
This delta function is only to be used in $d+1$ dimensions,
as we assume that only in such a situation
the regularization due to the extra dimensions
is taking place~\footnote{We are not able to show this in full generality,
and at this stage this rule is taken as an assumption.
One way to prove it explicitly 
would be to compute all integrals arising in perturbation theory
at arbitrary $d$ and check the location of the poles.}.
By using partial integration one casts the various loop 
integrals in a form which can be computed by sending first
$d\to 0$. At this stage one can use $\gamma^0=1$,
and no extra factors arise from the Dirac algebra in $d+1$ 
dimensions.  
This procedure will be exemplified in the subsequent calculations.
Having specified how to compute the ambiguous
Feynman graphs by continuation to $d+1$ dimensions
the DR scheme is now complete.

Now we are ready to perform the two-loop calculation in the ${\cal N}=1$ 
nonlinear sigma model using DR.
The bosonic vertices together with the ghosts, $V$  and $V_{DR}$ 
give the standard contribution, as for example
in \cite{Bastianelli:2002fv}.
The overall normalization of the fermionic path integral gives
the extra factor $2^{D\over 2}$  which equals the number of components 
of a Dirac fermion in a target space of even dimensions $D$.
This already produces the full expected result in (\ref{4.4.2}).

Thus the sum of the additional fermion graphs arising from
the cubic vertex contained in  
$\Delta S= \int_0^1 d\tau \,  {1\over 2\beta} \,
\dot x^\mu \omega_{\mu ab} \psi^a \psi^b$ 
and the contribution from the extra counterterm 
$ V_{DR}'$ 
must vanish at two loops.
The cubic vertex arise by evaluating the 
spin connection at the background point $x_0$ 
and reads  
$\Delta S_3= {1\over 2\beta} \omega_{\mu ab} 
\int_0^1 d\tau \,  \dot y^\mu 
\psi^a \psi^b$,
where  $y^\mu $ denotes the quantum fluctuations around the background 
point $x^\mu_0$ with vanishing boundary conditions at $\tau=0,1$.
Using Wick contractions (see appendix A.2 for the explicit form of the
bosonic propagators with vanishing Dirichlet boundary conditions)
we identify the following nontrivial contribution to $\la e^{-S^{int}}\ra$
(other graphs vanish trivially)
\bea
{1\over 2} \la (\Delta S_3)^2  \ra = 
\hspace{-.5cm}
\raisebox{-.78cm}{\scalebox{.7}{ 
{\includegraphics*[130pt,655pt][197pt,720pt]{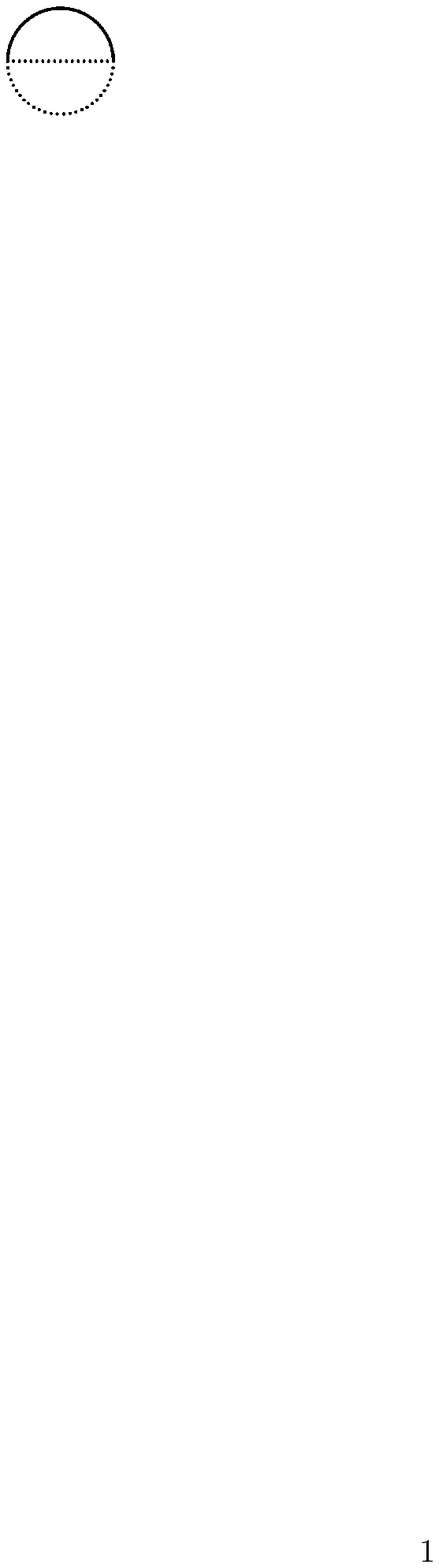}}}}
 = 
{1\over 2}
(-2) \Big ({1\over 2\beta}\omega_{\mu ab}\Big)^2 (-\beta^3)
\int_0^1 \!\! \! d\tau  \!\! \int_0^1 \!\! \! d\sigma \   
\ddeld (\tau, \sigma)
 [\Delta_{AF} (\tau,\sigma)]^2 
\label{4.4.13}
\eea
where dotted lines represent fermions. 
As usual, we denote with a left/right dot the derivative 
with respect to the first/second variable.
Using DR this contribution is regulated by
\bea
\int_0^1 \!\! \! d\tau  \!\! \int_0^1 \!\! \! d\sigma \
\ddeld (\tau, \sigma)
[ \Delta_{AF} (\tau, \sigma)]^2 \ \ 
\to \ \ -\int \!\! \int 
{ _\alpha{\Delta_\beta}(t,s)}\; {\rm tr}\, [\gamma^\alpha \Delta_{AF} (t,s) 
\gamma^\beta \Delta_{AF} (s,t)] 
\eea
where $  _\alpha{\Delta_\beta}(t,s) \equiv {\partial\over \partial t^\alpha} 
{\partial\over \partial s^\beta} \Delta(t,s) $
(note the minus sign obtained in exchanging $t$ and $s$
in the last propagator; it is the usual minus sign arising for fermionic
loops).
We can partially integrate $\partial_\alpha$ without picking
boundary terms and obtain
\bea
&&
 2 \int \!\! \int 
{ {\Delta_\beta}(t,s)} \; {\rm tr}\,  
[(\gamma^\alpha \partial_\alpha\Delta_{AF} (t,s)) 
\gamma^\beta \Delta_{AF} (s,t)]
\cr && 
=  2 \int \!\! \int 
{ {\Delta_\beta}(t,s)}\; {\rm tr}\,  [ \delta_A^{d+1}(t,s) 
\gamma^\beta \Delta_{AF} (s,t)]\cr
&&
= 2 \int  
{ {\Delta_\beta}(t,t)} \; {\rm tr}\,  [ 
\gamma^\beta \Delta_{AF} (t,t)]
\cr &&
 \to \ \ \ \- 2 
\int_0^1 \!\! \! d\tau \
{{\deld}(\tau,\tau)} \Delta_{AF} (0)  = 0
\label{dim-reg-1}
\eea
because $\Delta_{AF}(0) = {1\over 2} \epsilon (0) =0$ 
(and $\gamma^0=1$ at $d=0$).
As this example shows, the Dirac gamma matrices in $d+1$
dimensions 
are just a book-keeping device to keep track where 
one can use the Green equation ({\ref{4.4.11}).
Actually, the vanishing of this graph is achieved already before removing 
the regularization $d\to 0$ by using symmetric integration 
in the momentum space representation of $\Delta_{AF} (t,t)$.

Thus no contributions arise from the fermions at order
$\beta^2$, and this fixes 
\be 
V_{DR}' =0 \ .
\ee
This is exactly what one expects to preserve supersymmetry,
as the counterterm $V_{DR}$ is exactly canceled by the extra
potential term $V={1\over 8} R$ needed to have the correct
coupling to the scalar curvature in the Hamiltonian (\ref{4.4.ham}).
Thus dimensional regularization without any counterterm
preserves the supersymmetry of the classical ${\cal N}=1$ action
\bea 
S = {1\over \beta} \int_{0}^1 \!\! d\tau\, \Big [
{1\over 2}g_{\mu\nu} \dot x^\mu \dot x^\nu 
+ {1\over 2}  \psi_a (\dot \psi^a + \dot x^\mu \omega_\mu{}^a{}_b \psi^b)
\Big ] 
\eea
since the amount of the curvature coupling 
brought in by DR is of the exact amount to render 
the quantum Hamiltonian $\hat H$ supersymmetric.

To compare with TS, we can compute again the Feynman graph
(\ref{4.4.13}), but now using the TS rules. 
According to  \cite{deBoer:1995cb} we must use  
that $ \ddeld(\tau,\sigma)= 1-\delta(\tau,\sigma)$,
and integrate the delta function even if it acts on discontinuous functions.
The delta function is ineffective as $\epsilon(0) =0$, but the 
rest gives 
\bea
{1\over 2} \la (\Delta S_3)^2  \ra(TS) = {1\over 2}
(-2) \Big ({1\over 2\beta }\omega_{\mu ab}\Big)^2 (-\beta^3)  
\int_0^1 \!\! \! d\tau  \!\! \int_0^1 \!\! \! d\sigma \
{ 1\over 4}
={\beta\over 16}  (\omega_{\mu ab})^2  \ .
\label{4.4.TS}
\eea
This is canceled by using an extra counterterm
$V_{TS}'= {1\over 16} (\omega_{\mu ab})^2$
which at this order contributes with a term
$-\beta V_{TS}'$ evaluated at the background point $x_0$.
Thus, as expected, we recover the counterterm $V_{TS}'$
found in \cite{deBoer:1995cb}.

To summarize, we have proven that DR extended to fermions does not require
additional counterterms on top of those described in  
\cite{Bastianelli:2000nm}. In addition, supersymmetry requires 
that no counterterms should be added at all to the classical
sigma model action.

\subsection{Periodic boundary conditions}

We present here some comments on
the case of Majorana fermions with PBC.
The mode expansion of $\psi^a(\tau)$ has now only integer modes
\be
\psi^a
(\tau) = \sum_{n\in Z}
\psi^a_n 
\, {\rm e}^{2  \pi i n \tau} \ .
\label{4.4.17}
\ee
The zero modes $\psi^a_0 $
 of the free kinetic operator ($\partial_\tau$) are treated separately, 
and the unregulated propagator in the sector 
of periodic functions orthogonal to the zero mode reads 
\be
\la \psi'^a(\tau) \psi'^b(\sigma)\ra 
= \beta \delta^{ab} \Delta_{PF}(\tau-\sigma)\ ,
\ \ \ \ 
\Delta_{PF}(\tau-\sigma) =
\sum_{n\neq 0} 
{1\over 2 \pi i n }
 {\rm e}^{2  \pi i n (\tau-\sigma)} 
\label{4.4.18}
\ee
where $\psi'^a(\tau) =\psi^a(\tau)- \psi^a_0$.
The function $\Delta_{PF}$ satisfies
\be
\partial_\tau \Delta_{PF}(\tau-\sigma) =
\delta_{P}(\tau-\sigma)-1
\label{4.4.19}
\ee
with $\delta_{P}(\tau-\sigma)$ the Dirac's delta on periodic functions.
Its continuum limit can be obtained by summing up the Fourier series
and reads (for $(\tau -\sigma) \in [-1,1]$)
\be
\Delta_{PF}(\tau-\sigma) = {1\over 2}\epsilon (\tau -\sigma)
- (\tau -\sigma) \ .
\ee

The dimensionally regulated propagator is instead
\be
\la \psi'^a(t) \bar \psi'^b(s)\ra =
\beta\, \delta^{ab} \Delta_{PF}(t,s) 
\label{449}
\ee
where the function
\bea 
\Delta_{PF}(t,s)
= - i \int {d^d{\bf k}\over (2\pi)^d} 
\sum_{n\neq 0} 
 { 2 \pi n \gamma^0 
+ {\bf k \cdot \vec \gamma } \over (2 \pi n)^2+{\bf k}^2}\,
 {\rm e}^{2  \pi i n (\tau-\sigma)}
{\rm e}^{i{\bf k}\cdot ({\bf t}-{\bf s})}  
\label{4410}
\eea
satisfies
\be
\gamma^\alpha {\partial\over \partial t^\alpha} \Delta_{PF}(t,s)=
-{\partial\over \partial s^\beta} \Delta_{PF}(t,s) \gamma^\beta 
=(\delta_{P}(\tau-\sigma) -1)\delta^d({\bf t-s}) 
\ .
\label{4411}
\ee

Even if one uses PBC, one does not expect additional counterterms
in DR, as mentioned earlier. 
It could be interesting to check in DR 
the expected $\beta$-independence of the 
supertrace which computes the Witten index i.e. the chiral anomaly. 
This is given by the path integral with periodic boundary conditions
for both bosons and fermions 
\cite{Alvarez-Gaume:1983at,Alvarez-Gaume:1983ig,Windey:1983se}.
The treatment of the bosonic zero modes
is known to be somewhat delicate as a total derivative term may appear
at higher loops \cite{Bastianelli:2002fv,Hatzinikitas:1997md}.
However it should be possible to do a manifestly supersymmetric computation
using superfields, and one could thus check if these total derivative 
terms survive in the supersymmetric case and, in case they do, 
study their meaning 
\footnote{In a very recent paper, Kleinert and 
Chervyakov \cite{Kleinert:2003zq} have discussed how to avoid these total 
derivative terms which appear using naively the bosonic string inspired 
propagators.}.

\subsection{Curved indices}\label{ss:curved-indices}

It is interesting to consider as well the case of fermions
with curved target space indices.
This should be equivalent to the case of fermions with 
flat target space indices: it is just a change of integration
variables in the path integral. However it is an useful exercise
to work out, as some formulas will become simpler.
The classical ${\cal N}=1$ supersymmetric sigma model is now written as
\bea 
S = {1\over \beta}
\int_0^1 \!\! d\tau\, 
{1\over 2}g_{\mu\nu}(x) \Big [\dot x^\mu \dot x^\nu 
+ \psi^\mu [\dot \psi^\nu + \dot x^\lambda 
\Gamma_{\lambda\rho}^\nu(x) \psi^\rho] \Big ] \ . 
\label{4.5.1}
\eea 
The fermionic term could also be written more compactly using 
the covariant derivative ${D\over d\tau} \psi^\nu = \dot \psi^\nu 
+ \dot x^\lambda \Gamma_{\lambda\rho}^\nu(x) \psi^\rho$.
Note that the action is now expressed in terms of the metric 
and Christoffel connection only, and there is no need of introducing the 
vielbein and spin connection.

The treatment of the 
bosonic part goes on unchanged. For the fermionic part we can derive
the correct path integral measure by taking into account 
the jacobian for the change of variables from the free measure 
with flat indices 
\bea
D \psi^a = D(e^a{}_\mu(x)\psi^\mu) = 
{\rm Det}^{-1}(e^a{}_\mu(x)) D \psi^\mu = \left (
\prod_{0 \leq \tau<1} {1\over \sqrt{\det g_{\mu\nu}(x(\tau))}} \right )
 D \psi^\mu \ .
\eea
Note the inverse functional determinant appearing because of
the Grassmann nature of the integration variables.
This extra factor arising in the measure can be exponentiated
using bosonic ghosts $\alpha^\mu(\tau)$
with the same boundary condition of the fermions (ABC or PBC)
and it leads to the following extra term in the ghost action 
\bea
S_{gh}^{extra} = {1\over \beta} \int_0^1 \!\! d\tau\, 
{1\over 2}g_{\mu\nu}
(x) \alpha^\mu \alpha^\nu \ .
\eea
One can check that the counterterms of dimensional regularization 
are left unchanged. The full quantum action 
for the ${\cal N}=1$ supersymmetric sigma model now reads 
\bea 
S = {1\over \beta}
\int_0^1 \!\! d\tau\, 
{1\over 2}g_{\mu\nu}(x)\Big[\dot x^\mu \dot x^\nu + a^\mu a^\nu + b^\mu c^\nu 
+ \psi^\mu  [\dot \psi^\nu + \dot x^\lambda 
\Gamma_{\lambda \rho}^\nu(x) \psi^\rho] +
\alpha^\mu \alpha^\nu \Big ] 
\label{4.5.3}
\eea 
and
appears in the path integral as
\bea
Z=\int \!  Dx Da Db Dc D\psi D\alpha\  e^{- S} \ .
\eea

It is clear that supersymmetry is not broken by
the boundary conditions if one uses PBC. Then
the effects of the ghosts cancel by themselves: the ghosts have the
same boundary conditions and can be eliminated altogether 
from the path integral
\be
\left(\prod_{0\leq \tau<1}\sqrt{\det g_{\mu\nu}(x(\tau))} \right)
\left(\prod_{0\leq \tau<1}{1\over \sqrt{\det g_{\mu\nu}(x(\tau))}}\right)=1\ .
\ee
One can recognize that
the potential divergences arising in the bosonic $\dot x \dot x$ contractions
are canceled by the fermionic $\psi \dot\psi$ contractions, while
the remaining UV ambiguities are treated by dimensional regularization
as usual.
In this scheme it should be simpler for example to test
that the Witten index (i.e. the gravitational contribution to the chiral
anomaly for a spin 1/2 field) does not get higher order contributions
in worldline loops, and is thus $\beta$ independent.

If one uses ABC the ghosts have different boundary conditions. Hence
their cancellation is not complete, and they must be kept in the
action.

\section{Trace anomaly for a spin 1/2 field in 4 dimensions}

As a further test on the DR scheme applied to fermions, 
we compute the trace anomaly for a spin 1/2 fields in 4 dimension.
This anomaly is given by: 
$i)$
extending the formula (\ref{4.4.2}) to include the three-loop correction 
(order $\beta^2$ inside the round bracket), 
$ii)$
setting $D=4$,  
$iii)$
picking up the order $\beta^0$ term \cite{Bastianelli:1992be} and,
$iv)$
including an overall minus sign which takes care of the fermionic
nature of the target space loop.
The bosonic part has been computed already in DR
using Riemann normal coordinates
(see \cite{Bastianelli:2000dw}, use $\xi={1\over 4}$,
and recall our conventions on the scalar curvature
reported in appendix A.1). 
Multiplied by $2^{D\over 2}$ (the additional normalization
due the worldline ABC Majorana fermions) it reads 
\bea
Z_{bos} 
\!\!&=&\!\! 
{\rm tr}\, \la x_0| e^{-\beta \hat H}|x_0\ra_{bos}
\nonumber\\ 
&=&\!\! {2^{D\over 2}
\over (2 \pi \beta)^{D\over 2}} \Big (1 - {\beta\over 24} R 
+{\beta^2 \over 1152} R^2 + {\beta^2 \over 720} 
(R_{\mu\nu\lambda\rho}^2 -R_{\mu\nu}^2)
-{\beta^2 \over 480} \nabla^2 R 
+ O(\beta^3)\Big )  .
\label{tan}
\eea
We have now to include the fermionic contributions.
On top of Riemann normal coordinates we may use a Fock-Schwinger
gauge for the spin connection
$\omega_{\mu a b}(x_0+y) =  {1\over 2} y^\nu R_{\nu\mu ab}(x_0) + ...$
with $y^\mu$ the Riemann normal coordinates around the background
point $x^\mu_0$.
Then the leading quartic vertex $S_{4,f} =
{1\over 4 \beta} R_{\mu\nu ab} \int_{0}^1 d\tau\,    
y^\mu \dot y^\nu \psi^a \psi^b $ which
originates from the spin connection produces
the following 3-loop diagram 
\bea
\Big \la {1\over 2} (S_{4,f})^2 \Big \ra =
\hspace{-.5cm}
\raisebox{-.78cm}{\scalebox{.7}{ 
{\includegraphics*[130pt,655pt][197pt,720pt]{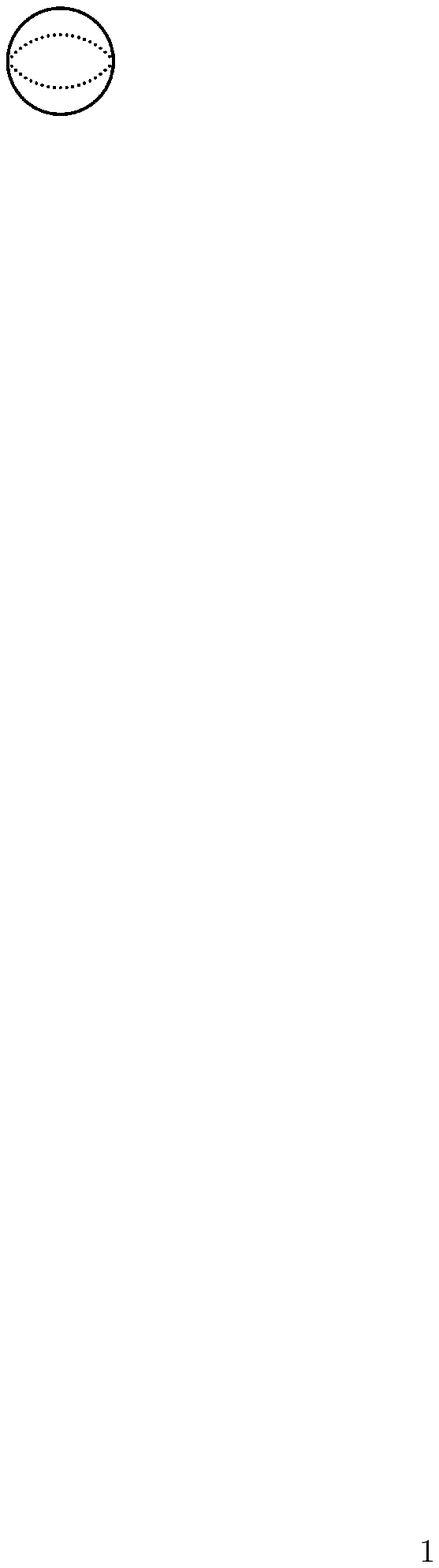}}}}
 = 
- {\beta^2\over 16} R^2_{\mu\nu ab}
\int_{0}^{1}  \!\!\! d\tau \! \int_{0}^{1}  \!\!\!  d\sigma \ 
( \ddeld\, \Delta - \ddel \deld )\, \Delta_{AF}^2
\label{8.2.18}
  \eea
where all functions $\Delta $ and $\Delta_{AF}$ 
are functions of $\tau$ and $\sigma$ in this precise order (recall that 
$\Delta_{AF}$ is antisymmetric, 
$\Delta_{AF}(\tau,\sigma) = - \Delta_{AF}(\sigma,\tau)$). 

Now we regulate this graph in DR as follows
(the second contribution in (\ref{8.2.18})
does not need regularization
and could be directly computed at $d=0$, but we carry it along anyway) 
\bea
&&\hskip-.75cm \int_{0}^{1}  \!\!\! d\tau \! \int_{0}^{1}  \!\!\!  d\sigma \ 
(\ddeld\, \Delta - \ddel \deld )\, \Delta_{AF}^2
\  \to  \ 
\int \!\! \int 
( _\alpha{\Delta_\beta} \  \Delta
 - { _\alpha{\Delta}}\, \Delta_\beta  )\, {\rm tr}[-
\gamma^\alpha\Delta_{AF}(t,s) \gamma^\beta\Delta_{AF}(s,t)] 
\nonumber
\\[1mm] &&\hskip-.75cm
=
\int \!\! \int 
 { {\Delta_\beta}}\  \Delta \,  
{\rm tr}[2 
\underbrace{(\gamma^\alpha {\partial \over \partial t^\alpha}
\Delta_{AF}(t,s))}_{\delta_A^{d+1}(t-s)} 
\gamma^\beta\Delta_{AF}(s,t)] 
- 2
\int \!\! \int 
{ _\alpha{\Delta}}\, \Delta_\beta  \, {\rm tr}[-
\gamma^\alpha\Delta_{AF}(t,s) \gamma^\beta\Delta_{AF}(s,t)] 
\cr 
&&\hskip-.75cm
= 0 - 2 
\int \!\! \int 
{ _\alpha{\Delta}}\, \Delta_\beta  \, {\rm tr}[-
\gamma^\alpha\Delta_{AF}(t,s) \gamma^\beta\Delta_{AF}(s,t)] 
\nonumber
\\[1mm] 
&&\hskip-.75cm 
\to \ \ 
-2 \int_{0}^{1}  \!\!\! d\tau \! \int_{0}^{1}  \!\!\!  d\sigma \ 
 \ddel \deld  \Delta_{AF}^2 =
-2 \int_{0}^{1}  \!\!\! d\tau \! \int_{0}^{1}  \!\!\!  d\sigma \ 
 \ddel \deld  {1\over 4} =-2 (-{1\over 12})({1\over 4})={1\over 24}
\label{dim-reg-2}
\eea 
where we have integrated by parts the $\alpha$ derivative
in  $ _\alpha{\Delta_\beta} $, which  then produces a delta function 
when acting on fermions (``equations of motion terms'').
This delta function is 
integrated in $d+1$ dimensions and  gives a vanishing contribution
since $\Delta_{AF}(0)= 0$. The remaining terms are then computed 
at $d\to 0$.
Thus 
\be
\Big \la {1\over 2} (S_{4,f})^2 \Big \ra =
- {\beta^2\over 384} R^2_{\mu\nu ab} \ .
\ee

This  fermionic contribution must now be added 
to the terms inside the round bracket of eq. (\ref{tan}). 
Setting $D=4$ 
one recognizes the following anomaly
\bea
Z|_{\beta^0-term}
={ 1 \over 4 \pi^2} \Big (
{1 \over 288} R^2 - {7 \over 1440} R_{\mu\nu\lambda\rho}^2 
-{1 \over 180} R_{\mu\nu}^2 -{1 \over 120} \nabla^2 R \Big ) \ .
\label{tan2}
\eea
This is the correct trace anomaly for a Dirac fermion in 4 dimensions 
once we include the minus sign due to the target space fermionic loop.

\section{One-loop effective action for a Dirac field in a   \newline
             gravitational  background}

It is known that, for a wide class of field theories, the one-loop
effective action and the relative $N$-point vertex functions can be computed 
using one-dimensional path integrals \cite{Strassler,Schubert:2001he}. 
Two of us have presented in \cite{Bastianelli:2002fv}
the extension of this formalism to include a gravitational background,
considering the simplest case of a scalar field. 
The extension of DR to worldline fermions allows us to do the same 
for a Dirac field.
We will get an expression for the effective action from which
we derive explicitly the one- and two-point correlation functions, 
namely the contribution to the tadpole and self-energy of the graviton. 
We perform this program
considering both flat and curved indices for the worldline Majorana
fermions. The use of flat indices produces an effective action 
$\bar\Gamma[e_a{}_\mu]$ which is naturally a functional of the vielbein.
The use of curved indices produces instead an effective action 
$\Gamma[g_{\mu\nu}]$ which is naturally a functional of the metric.
Local Lorentz invariance guarantees that 
$\bar\Gamma[e_a{}_\mu] = \Gamma[g_{\mu\nu}(e_a{}_\mu)]$ .
In the following we shall discuss both cases.
As we shall see, the simplest set up is to use curved indices:
in this case the sigma model couples linearly
to the metric fluctuations $h_{\mu\nu} = g_{\mu\nu}- \delta_{\mu\nu}$, 
and the effective $N$-point vertices for the metric
are obtained by integrating over the proper time the quantum average 
of $N$ graviton vertex operators.

\subsection{The worldline formalism}   
\label{ss:world-line}

Let us consider the one-loop effective action obtained by quantizing a Dirac 
field $\Psi$ coupled to gravity through the vielbein $e_a{}_\mu$  
\bea
S[\Psi,\bar \Psi, e_a{}_\mu ]
= \int \!d^Dx \; e\, \bar{\Psi} (\rldd + m)\Psi  
\label{uno}
\eea
where $ e= \det e^a{}_\mu$, $\omega_{\mu ab} $ is the spin connection, and

\be
\rldd = \gamma^a e_a{}^\mu \nabla_{\!\mu} \ , \quad \quad 
\nabla_{\!\mu} = \partial_\mu + {1\over 4}\omega_{\mu ab} \gamma^a \gamma^b\ . 
\ee
The effective action depends on the background vielbein field
$e_a{}_\mu$  and formally reads as  
($ e^{-\bar\Gamma[e_a{}_\mu]} \equiv \int {\cal D}\Psi\, {\cal D}\bar \Psi 
\ e^{-S[\Psi,\bar\Psi, e_a{}_\mu]} = {\rm Det} (\rldd + m)$) 
\be
\bar\Gamma[e_a{}_\mu] = -\log {\rm Det} (\rldd + m) \ .
\ee
For a Dirac field one does not expect anomalies (the euclidean effective
action is real) and one can exploit standard arguments to write
\bea
\bar\Gamma[e_a{}_\mu] 
\!\! &=& \!\!
 -\log [{\rm Det} (\rldd + m) {\rm Det} (-\rldd + m)]^{1\over 2} 
= - {1\over 2} {\rm Tr} \log  (-\rldd^2 + m^2) \cr
\!\! &=& \!\!
 - {1\over 2} {\rm Tr} \log  \Big (-\nabla^2 + {1\over 4}R+m^2\Big ) 
\ .
\label{trace}
\eea
In this formula we
recognize the logarithm of an operator which up the mass term
is proportional to the supersymmetric hamiltonian (\ref{4.4.ham}).
Thus, we can immediately write down a path integral representation
for the effective action in terms of a proper time as
\bea
\bar\Gamma[e_a{}_\mu] 
= {1\over 2} \int_0^\infty {dT\over T } \oint_{PBC} \!\!\!\!\!\!\!
{\cal D}x^{\mu}
\oint_{ABC} \!\!\!\!\!\!\!
D\psi^a \; e^{-S[x^\mu,\psi^a; e_a{}_\mu]}                 
\label{effaction}
\eea
where \footnote{Presumably, this final action can be obtained also by gauge 
fixing the locally supersymmetric formulation of the spinning particle
action \cite{Henneaux:ma,Henty:hh}, at least in the massless case,
as the corresponding ghosts decouple from the 
background geometry and can be ignored.}
\bea
S[x^\mu,\psi^a; e_{a\mu}]  =
\int_{0}^{1} \! \! d\tau\, \Big [
{1\over 4 T} g_{\mu\nu}(x) \dot x^\mu \dot x^\nu + \frac{1}{4 T} 
\psi_a\dot{\psi}^a+
 \frac{1}{ 4T}\dot x^{\mu}\w_{\mu ab}(x)\, \psi^a\psi^b +T m^2 
\Big ] \ .
\label{gfaction}
\eea
The subscripts PBC and ABC remind of the boundary condition at $\tau=0,1$,
periodic for the bosonic coordinates $x^\mu(\tau)$ and antiperiodic for the
fermionic ones $\psi^a(\tau)$: these boundary conditions have to be imposed 
to obtain the trace in (\ref{trace}).
We have used a rescaled proper time $T={\beta\over 2}$ with respect 
to the previous sections to agree with standard
normalizations used in the worldline formalism~\cite{Schubert:2001he}.
We have not added any counterterm since we are going to use dimensional
regularization to compute the path integral~\footnote{
Let us recall that other regularization schemes 
(such as time slicing \cite{deBoer:1995cb}) 
require  additional non-covariant counterterms.}. 
Of course, the covariant measure in (\ref{effaction}) contains the ghost 
fields 
\bea
{\cal D}x^{\mu} = Dx^{\mu} \prod_{ 0 \leq \tau < 1}
 \sqrt{\det g_{\mu\nu}(x(\tau))} = Dx^{\mu} \oint_{PBC}  \!\!\!\!\!\!\!
{ D} a^{\mu} { D} b^{\mu}
 { D} c^{\mu} \;\; {\rm e}^{- S_{gh}[x,a,b,c]}                       
\label{xmeasure}
\eea
where
\bea
S_{gh}[x,a,b,c]
 = \int_{0}^{1} \!\! d\tau \, {1\over 4T}g_{\mu\nu}(x)(a^\mu a^\nu
 + b^\mu c^\nu) \ .
\eea

One may also compute the effective action directly as a functional 
of the metric.
This is achieved by using the sigma model written in terms of the
Majorana fermions with curved indices. The corresponding formula is
\bea
\Gamma[g_{\mu\nu}] 
= {1\over 2} \int_0^\infty {dT\over T } \oint_{PBC} \!\!\!\!\!\!\!
{\cal D}x^{\mu}
\oint_{ABC} \!\!\!\!\!\!\!
{\cal D}\psi^\mu \; e^{-S[x^\mu,\psi^\mu; g_{\mu\nu}]}                 
\label{effactioncur}
\eea
with
\bea
S[x^\mu,\psi^\mu; g_{\mu\nu}]  =
\int_{0}^{1} \! \! d\tau\, \Big [
{1\over 4 T} g_{\mu\nu}(x) (\dot x^\mu \dot x^\nu + 
\psi^\mu \dot{\psi}^\nu +\psi^\mu
\Gamma_{\lambda\rho}^\nu(x) \dot x^\lambda \psi^\rho )
+T m^2 \Big ] \ .
\label{gfactioncur}
\eea
Note that the covariant fermionic measure now contains the new 
bosonic ghost $\alpha^\mu$ 
\bea
{\cal D}\psi^{\mu} = D\psi^{\mu} 
\prod_{0\leq\tau<1}{1\over \sqrt{\det g_{\mu\nu}(x(\tau))}}
= D\psi^{\mu} \oint_{ABC}  \!\!\!\!\!\!\!
{ D} \alpha^{\mu} 
\;\; {\rm e}^{- S^{extra}_{gh}[x,\alpha]}                       
\label{xmeasurecur}
\eea
with
\bea
S^{extra}_{gh}[x,\alpha] = \int_{0}^{1} 
\!\! d\tau\, {1\over 4T}g_{\mu\nu}(x)\alpha^\mu \alpha^\nu \ .
\eea
The fermionic term in the action (\ref{gfactioncur})
may be written using the covariant derivative as 
$g_{\mu\nu} \psi^\mu {D\over d\tau}{\psi}^\nu$,
making manifest its geometrical meaning. However, 
one can write the Christoffel connection directly in terms of the metric
and, because of the Grassmannian nature of the fields $\psi^\mu$,
the action simplifies to
\bea
S[x^\mu,\psi^\mu; g_{\mu\nu}] =
\int_{0}^{1} \! \! d\tau\, \Big [
{1\over 4 T} g_{\mu\nu}(x) (\dot x^\mu \dot x^\nu + 
\psi^\mu \dot{\psi}^\nu) -  \frac{1}{ 4T} \partial_\mu g_{\nu\lambda}(x)
\psi^\mu\psi^\nu \dot x^{\lambda} +T m^2 \Big ]
\label{gfactioncur2}
\eea
which shows that there is only a linear coupling to 
the background $g_{\mu\nu}(x)$.
To summarize, we have two options for representing the effective action 
in the worldline formalism, and we will consider both of them. 

The next step is to discuss how to treat the boundary conditions. 
Due to the translational invariance of the resulting propagators, 
we adopt the ``string inspired'' option:
one expands the coordinate fields with periodic
boundary conditions into Fourier modes 
and then separates the zero mode $x^\mu_0=\int^1_0d\tau \, x^\mu(\tau)$
from the quantum fluctuations $y^\mu(\tau)= x^\mu(\tau)-x^\mu_0$.
The latter have an invertible kinetic term and the
integration over the constants zero mode $x_0^\mu$ 
is performed separately. 
For the alternative option of using Dirichlet boundary conditions,
see a discussion in \cite{Bastianelli:2002fv}.
Other options for treating the zero modes can be found in
\cite{FT,Schubert:2001he}.

These subtleties do not arise for
the anticommuting variables $\psi^a$ as the boundary conditions 
are now antiperiodic and  the kinetic term has no zero mode.
All these propagators are collected in section A.3 of the appendix.

For later convenience, it is useful to introduce the following notations
\bea
&& 
\bar{\Gamma}_{(x_1,\ldots,x_N)}^{a_1\nu_1\cdots a_N\nu_N} =
\frac{\delta^N\bar\Gamma}{\delta e_{a_1\nu_1}(x_1)\cdots\,
\delta e_{a_N\nu_N}(x_N)}\Big|_{e_{a \nu}=\delta_{a \nu}}                  
\label{bargamma-N}
\\
&& 
\Gamma_{(x_1,\ldots,x_N)}^{\mu_1\nu_1\cdots\mu_N\nu_N} =
\frac{\delta^N\Gamma}{\delta g_{\mu_1\nu_1}(x_1)\cdots\,
\delta g_{\mu_N\nu_N(x_N)}}\Big|_{g_{\mu\nu}=\delta_{\mu\nu}}           
\label{gamma-N}
\eea
and the corresponding Fourier transform for the vielbein 
vertex functions
\bea\label{notation1}
\tilde {\bar \Gamma}_{(p_1,..., p_N)}^{a_1\nu_1\cdots a_N\nu_N} 
\! &=& (2\pi)^D
 \delta^D(p_1+\cdots+p_N)\bar \Gamma_{(p_1,\dots, p_N)}^{a_1\nu_1\cdots 
a_N\nu_N} 
\nonumber\\
&=&
 \int dx_1\cdots dx_N\, e^{ip_1\cdot x_1 +\cdots+ip_N\cdot x_N} \,
 \bar \Gamma_{(x_1,\dots, x_N)}^{a_1\nu_1\cdots a_N\nu_N}  \quad
\eea
plus a similar one for the metric vertex functions.
For $N=1,2$,  and using 
$\bar\Gamma[e_a{}_\rho] = \Gamma[g_{\mu\nu}(e_a{}_\rho)]$ 
together with the relation
$g_{\mu\nu} = \delta_{ab} e^a{}_\mu e^b{}_\nu$,
one finds 
\bea
&&\bar{\Gamma}_{(x)}^{\mu\nu}=2\Gamma_{(x)}^{\mu\nu} 
\label{onepoint}\\
&&\bar{\Gamma}_{(x,y)}^{\mu_1\nu_1\mu_2\nu_2}=               \label{twopoint}
4\Gamma_{(x,y)}^{\mu_1\nu_1 \mu_2\nu_2}
+2\Gamma_{(x)}^{\nu_1\nu_2}\,
\delta^{\mu_1\mu_2}\,\delta^D(x-y)~.
\label{gamma-bargamma}
\eea

Following a standard technique, one can obtain the vertex functions directly 
in momentum space \cite{Schubert:2001he}. 
Let us describe it for the effective action 
 $\bar\Gamma[e_a{}_\nu]$.
One considers $\bar\Gamma[e_a{}_\nu]$ as a power series in
$c_{a\nu} \equiv e_{a\nu}-\delta_{a\nu}$ 
(note that this definition induces a relative expression for the metric: 
$g_{\mu\nu} =  \delta_{\mu\nu}+ c_{\mu\nu}+
c_{\nu\mu}+ c_{a \mu}c_{\nu}^a$, where $c_{\mu\nu}=c_{a\nu}\delta_\mu^a$),
takes the $c^N$ term as a sum of $N$ plane waves of given 
polarizations (our polarization tensors include the gravitational 
coupling constant)
\bea
c_{a \nu}(x) = \sum_{i=1}^{N} \varepsilon_{a \nu}^{(i)} e^{ip_i \cdot x} \ ,
\label{quindici}
\eea
and then picks up the terms linear in each $\varepsilon^{(i)}_{a \nu} $:
this gives directly the $N$-point function in momentum space
\bea
\tilde
{\bar\Gamma}^{\varepsilon_1\cdots\varepsilon_N}_{(p_1,\ldots,p_N)} 
\equiv  
\varepsilon^{(1)}_{a_1\nu_1} \cdots \varepsilon^{(N)}_{a_N\nu_N}  \,
 \tilde {\bar\Gamma}^{a_1\nu_1\cdots a_N\nu_N}_{(p_1,\ldots,p_N)} 
\eea
(the tilde symbol  can be dropped by removing the 
momentum delta functions as in (\ref{notation1})).

In the following sections we are going to compute the one- and two-point 
correlation functions.
We will employ the worldline ``string inspired'' propagators together 
with dimensional regularization on the worldline (and in target space).

\subsection{One- and two-point functions from $\bar\Gamma[e_a{}_\nu] $} 

The one-point vertex function can be depicted by the Feynman diagram of 
fig. 1 where the external line refers to the vielbein.
It gives the Dirac particle contribution to the cosmological constant.
The recipe just outlined tells that the term in the effective action linear 
in $c_{a \nu}$, and with $c_{a \nu}$
expressed as a single plane wave, produces
\bea
\tilde{\bar{\Gamma}}^{\varepsilon}_{(p)} \!\! &=& \!\!
{1\over 2} \int_{0}^{\infty}
{dT\over T} \; e^{-m^2 T} { 2^{D\over 2} \over (4 \pi T)^{D\over 2} }
\int d^Dx_0 \: \left(-{1\over 4 T }\right)                
\nonumber\\
&\times& \int_0^1 \!\! d\tau \, \biggl \{ \biggl \la 2 \varepsilon_{(\mu \nu)}
(\dot y^{\mu} \dot y^\nu +
a^{\mu} a^\nu + b^{\mu} c^\nu) e^{ip \cdot (x_0 + y)} \biggr \ra +
\la \dot y^{\mu}  \w_{\mu ab}^{(1)}(x_0+y) \ra
\la \psi^a \psi^b \ra \biggr \} \quad  
\eea
where the superscript on $\w_{\mu ab}^{(1)}$ denotes the part linear in
$\varepsilon_{a \nu}$, and round brackets around indices denote
symmetrization normalized to one.

\begin{figure}[ht]
\centering
\includegraphics{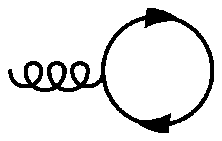}
\caption{\em Graviton tadpole.}
\label{fig3}
\end{figure}

\noindent
It can be immediately noted that the contribution of the spin connection 
term vanishes,
being proportional to $\omega_{\mu ab} \delta^{ab} \Delta_{AF}(0)=0$.
Therefore everything proceeds as in the scalar field case  
\cite{Bastianelli:2002fv}, and the one-point function reads
\bea
\bar\Gamma^{\mu\nu}_{(0)} = 2^{D \over 2}
{\delta^{\mu\nu} \over 2} 
{(m^2)^{D\over 2}\over (4 \pi)^{D\over 2} }
\Gamma\Big(-{D\over 2}\Big)  \ .
\label{1ptf}
\eea
Clearly it diverges for even target space dimension $D$ and 
renormalization is needed.

Let us now discuss the two-point vertex function. 
We set
\bea
c_{a \nu}(x) = \varepsilon_{a \nu}^{(1)} e^{ip_1 \cdot x}
 + \varepsilon_{a \nu}^{(2)} e^{ip_2 \cdot x}    \ .
 \label{twowaves}
\eea
One sees that there are two kinds of contributions, illustrated by 
the Feynman graphs in figures \ref{fig5} and \ref{fig4}, 
which we denote as $\Delta_1\bar\Gamma^{\mu\nu\alpha\beta}$ and 
$\Delta_2\bar\Gamma^{\mu\nu\alpha\beta}$, respectively.

\begin{figure}[hb]
\centering
\includegraphics{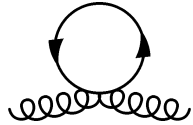}
\caption{\em One-vertex graph for graviton self-energy.}
\label{fig5}
\end{figure}

In the first one there is just one vertex. It is simple to compute it,
being quite similar to the tadpole. It reads
\bea
\Delta_1 \tilde{\bar \Gamma}_{(p_1,p_2)}^{\varepsilon_1 \varepsilon_2} 
\!\! &=&\!\!
{1\over 2} \int_{0}^{\infty}
{dT\over T} \, e^{-m^2 T} { 2^{D\over 2} \over (4 \pi T)^{D\over 2} }
\int d^Dx_0 \, \left(-{1\over 4 T}\right)  \nonumber\\
&\times& \!\! \int_0^1 \!\! d\tau\,
\biggl \{ \biggl \la c_{a \mu}c^a_{\phantom{a}\nu}  ( \dot y^{\mu} \dot y^{\nu}
+a^{\mu} a^\nu + b^{\mu} c^\nu) \biggr \ra +
\la \dot y^{\mu} \w_{\mu ab}^{(2)}(x_0+y) \ra
\la \psi^a  \psi^b \ra \biggr \}\Bigg|_{{\rm m.l.}}
\eea
where $\w_{\mu ab}^{(2)}$ is the part of the spin connection
quadratic in the $c_{a\mu}$ field; 
the prescription m.l. (multi-linear) refers to
the two different polarization tensors.
The contribution from the spin connection term vanishes for the same reason
as before. We are then left with the bosonic contribution, which gives 
\bea
\Delta_1\bar\Gamma^{\mu\nu\alpha\beta}_{(p,-p)}=
{2^{D \over 2} \over 2}\, 
\delta^{\mu\alpha}\delta^{\nu\beta}
{(m^2)^{D\over 2}\over (4 \pi)^{D\over 2} } \Gamma\Big(-{D\over 2}\Big)
\label{2tad}
\eea
where, according to the notation (\ref{notation1}),
we have factored out $(2\pi)^D \delta^D(p_1+p_2)$ and used $p=p_1=-p_2$.

The two-vertex graph of fig.~\ref{fig4} produces
\begin{figure}[ht]
\centering
\includegraphics{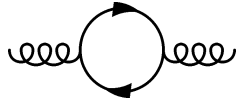}
\caption{\em Two-vertex graph for graviton self-energy.}
\label{fig4}
\end{figure}

\bea
\Delta_2 \tilde{\bar \Gamma}_{(p_1,p_2)}^{\varepsilon_1 \varepsilon_2} 
\!\! &=&\!\!
{1\over 2} \int_{0}^{\infty} {dT\over T} \, e^{-m^2 T} {2^{D\over 2} 
 \over  (4 \pi T)^{D\over 2} }\nonumber\\ 
&\times&\!\!
\int \! d^Dx_0\, \biggl \la {1 \over 2 }
  \left (\int_0^1 \! \!d\tau \, \Big [ {1\over 2 T}
  c_{(\mu\nu)} ( \dot y^{\mu} \dot y^{\nu} + a^{\mu} a^{\nu} + b^{\mu} c^{\nu})
  + {1\over 4T} \dot y^{\mu}\w^{(1)}_{\mu ab} \psi^a \psi^b \Big] 
\right )^{\!\!2}
  \, \biggr \ra  \biggr|_{\rm m.l.}~.                      \nonumber
\eea
Three kinds of contributions are included in the previous expression: 
$i)$ the square of the bosonic part which yields a term proportional
to the contribution of a scalar field 
(non-minimally coupled with $\xi =1/4$, 
already computed in \cite{Bastianelli:2002fv});
 $ii)$ the mixed terms of the product which are
zero, again being proportional to $\omega_{\mu ab} \delta^{ab} 
\Delta_{AF}(0)$; $iii)$ the square of the fermionic term 
which contains
\bea
\omega^{(1)}_{\mu ab}(x_0+y) = \sum_{i=1}^2\, 
\Big (-i p_\mu \varepsilon^{(i)}_{[ab]} +i\varepsilon^{(i)}_{\mu [a} 
p_{b]}-ip_{[a}\varepsilon^{(i)}_{b]\mu} \Big)\, 
e^{ip_i\cdot (x_0+y)}
\eea
where square brackets around indices denote antisymmetrization 
normalized to one. This third term produces the following contribution
\bea
&&\int_{0}^{\infty} {dT\over 32T^3} \, e^{-m^2 T} {2^{D\over 2} 
 \over  (4 \pi T)^{D\over 2} }
\Big (-i p_\mu \varepsilon^{(1)}_{[ab]}
+i\varepsilon^{(1)}_{\mu [a} p_{b]}-ip_{[a}\varepsilon^{(1)}_{b]\mu}\Big )
\Big (-i p_\rho \varepsilon^{(2)}_{[cd]}
+i\varepsilon^{(2)}_{\rho [c} p_{d]}-ip_{[c}\varepsilon^{(2)}_{d]\rho}\Big)
\nonumber\\
&&\hskip1.6cm
\times\int_0^1 \!\! d\tau \! \int_0^1 \!\! d\sigma
\, \left\la e^{ip\cdot y} \dot y^\mu 
\psi^a\psi^b(\tau) \, e^{-ip\cdot y} \dot y^\rho 
\psi^c\psi^d(\sigma)\right\ra \ .
\label{fermi-contribution}
\eea
After performing Wick contractions, the second line 
of this expression becomes 
\bea
&&(\delta^{ac}\delta^{bd}-\delta^{ad}\delta^{bc}) (2T)^3\nonumber\\
&&\times
\int_0^1 \!\! d\tau \! \int_0^1 \!\! d\sigma\,
\left\{\delta^{\mu\rho}\,\ddeld(\tau-\sigma)
  +2T p^\mu p^\rho \, \ddel^2(\tau-\sigma) \right\}
  e^{-2 T p^2 \Delta_0(\tau-\sigma)} \Delta_{AF}^2(\tau-\sigma)~,
\eea
where $\Delta_0(\tau-\sigma) = \Delta(\tau-\sigma) -\Delta(0)$, and 
needs worldline regularization.
Following the rules of dimensional regularization we write the last line 
of the above expression as we would have done starting from the action 
in $1+d$ dimensions
\bea
&& \int \!\! \int   \left\{\delta^{\mu\rho}
  {}_{\alpha}\!\!\:\Delta_{\beta}(t-s) +2T p^\mu p^\rho {}_{\alpha}\!\!\:
  \Delta(t-s)\,{}_{\beta}\!\!\:\Delta(t-s) \right\} e^{-2 T p^2 \Delta_0(t-s)} 
 \nonumber\\   && \times \;
  {\rm tr}\left[\gamma^{\alpha}\Delta_{AF}(t-s)\gamma^{\beta}
\Delta_{AF}(t-s)\right]
\label{66}
\eea
and perform an integration by parts on the $\alpha$ index
of the first term, as already explained in~(\ref{dim-reg-1}) and 
(\ref{dim-reg-2}), to get the following result
\bea
-{1\over 2}T p^2 \left(\delta^{\mu\rho}-{p^\mu p^\rho\over p^2}\right)
 \int_0^1 \!\! d\tau \left(\tau-{1\over 2}\right)^{\!\!2}
 e^{-T p^2 (\tau-\tau^2)} \ .
\eea
The remaining worldline integral can be computed
as described in section A.4 of the appendix.
Using the result into eq.~(\ref{fermi-contribution}) gives
\bea
\varepsilon^{(1)}_{\mu\nu}\varepsilon^{(2)}_{\alpha\beta}\,
{1\over 8}{2^{D \over 2}\over (4 \pi )^{D\over 2}}\,
\Gamma\Big(1-{D\over 2}\Big)\, p^2\,
  \left((P^2)^{{D\over 2}-1} -(m^2)^{{D\over 2}-1} \right) \, S_2^{\mu\nu
\alpha\beta}
\eea
where $P^2$ and $S_2$ are defined below. Collecting all terms, we find
for the two-vertex part of the self-energy  
\bea
\Delta_2 \bar\Gamma_{(p,-p)} \!\! &=&\!\! 
{ 2^{{D \over 2}}\over 2(4 \pi )^{D\over 2}}\left\{\Gamma\Big(-{D\over 2}\Big)
  \biggl [ (m^2)^{{D\over 2}} (R_1-R_2) 
  + \left((P^2)^{{D\over 2}} -(m^2)^{{D\over 2}}\right) (S_1  +S_2)
  \biggr]\right.\nonumber\\
&+& \!\! \left.{1\over 4} \Gamma\Big(1-{D\over 2}\Big)\, p^2\,
  (P^2)^{{D\over 2}-1}S_2\right\}~.
\label{xx}
\eea
Consequently, the full graviton self-energy 
obtained by summing (\ref{2tad}) and (\ref{xx}) reads
\bea
\bar\Gamma_{(p,-p)} \!\! &=&\!\!
{ 2^{{D \over 2}}\over 2 (4 \pi )^{D\over 2}}\left\{
\Gamma\Big(-{D\over 2}\Big)
  \biggl [ (m^2)^{{D\over 2}} (R_1-{1\over 2}R_2+{1\over 2}\bar R_2-S_1-S_2) 
  + (P^2)^{{D\over 2}}(S_1  +S_2)
  \biggr]\right.\nonumber\\
&+&\!\! \left.{1\over 4} \Gamma\Big(1-{D\over 2}\Big)\, p^2\,
  (P^2)^{{D\over 2}-1}S_2\right\}
\label{2ptf}
\eea
where, as in \cite{Bastianelli:2002fv}, we have suppressed
tensor indices, and used the following basis of dimensionless tensors 
symmetric in each pair of indices, $(\mu,\nu)$ and $(\alpha,\beta)$, and symmetric 
under the exchange of the two pairs
\bea
R_1^{\mu\nu\alpha\beta} \!\! &=& \!
\delta^{\mu\nu}\delta^{\alpha\beta}
\nonumber \\
R_2^{\mu\nu\alpha\beta} \!\! &=& \!
\delta^{\mu\alpha}\delta^{\nu\beta}+ \delta^{\mu\beta}\delta^{\nu\alpha} 
\nonumber \\
R_3^{\mu\nu\alpha\beta} \!\! &=& \!
{1\over p^2} \, (\delta^{\mu\alpha} p^\nu p^\beta +
\delta^{\nu\alpha} p^\mu p^\beta +
\delta^{\mu\beta} p^\nu p^\alpha +
\delta^{\nu\beta} p^\mu p^\alpha ) \nonumber\\
R_4^{\mu\nu\alpha\beta} \!\! &=& \!
 {1\over p^2} \, (\delta^{\mu\nu} p^\alpha p^\beta
+\delta^{\alpha\beta} p^\mu p^\nu)
\nonumber \\
R_5^{\mu\nu\alpha\beta} \!\! &=& \!
 {1\over p^4}\, p^\mu p^\nu p^\alpha p^\beta~,
\label{r-tensors}
\eea
along with the tensor~\footnote{We need such a tensor because the
  expression~(\ref{2tad}), and hence~(\ref{2ptf}), does not respect the
  symmetries of the $R$ tensors in~(\ref{r-tensors}). This is consistent with
  the local Lorentz Ward identities, see appendix~\ref{section:wi}.}
\bea
\bar R_2^{\mu\nu\alpha\beta} \!\! &=& \!
\delta^{\mu\alpha}\delta^{\nu\beta}- \delta^{\mu\beta}\delta^{\nu\alpha}~.
\eea
For simplicity we have introduced the manifestly transverse combinations
\bea
S_1^{\mu\nu\alpha\beta} \!\! &=& \!\!
R_1 -  R_4 + R_5 = 
\left(\delta^{\mu\nu}-{p^\mu p^\nu\over p^2}\right)
\left(\delta^{\alpha\beta}-{p^\alpha p^\beta\over p^2}\right)\\
S_2^{\mu\nu\alpha\beta} \!\! &=& \!\!
R_2 -  R_3 + 2 R_5 = 
2 \left(
\delta^{\underline{\mu}\overline{\alpha}}-{p^{\underline{\mu}}
p^{\overline{\alpha}}\over p^2}\right)\left(\delta^{\underline{\nu}
\overline{\beta}}-{p^{\underline{\nu}}p^{\overline{\beta}}\over p^2}\right)
\eea
and defined
\bea
(P^2)^x = \int_0^1 d\tau\, (m^2 + p^2 (\tau -\tau^2))^x~.
\eea
Further details may be found in section A.4 of the appendix. 

The final results for the one- and two-point functions,
eqs. (\ref{1ptf}) and (\ref{2ptf}), satisfy the gravitational Ward 
identities (see appendix A.5). Of course, one may now 
extract the divergent part
and renormalize these functions in the chosen spacetime
dimensions
\footnote{If one is interested in odd dimensions, then there
is no divergence at one loop, but the formulas should be
modified by substituting
$ 2^{{D \over 2}} \to 2^{[{D \over 2}]}$
to account for the correct number of components of a Dirac spinor.
Here $[{D \over 2}] $ denotes the integer part of ${D \over 2} $.}.

\subsection{One- and two-point functions from $\Gamma[g_{\mu\nu}] $}  

In this section we describe the calculation of the one- and two-point 
functions employing curved indices for the worldline Majorana fermions.
As already explained in sections~\ref{ss:curved-indices}
and~\ref{ss:world-line}, the total action including the ghost fields is given
 by
\bea
S &=& \int_{0}^{1}\! \!d\tau\, \left\{ {1\over 4 T} g_{\mu\nu}(x)
  (\dot x^\mu \dot x^\nu + a^\mu a^\nu +b^\mu c^\nu +\psi^\mu \dot\psi^\nu
  +\a^\mu \a^\nu)\right.\nonumber\\
&&\hskip1.5cm \left.
- \frac{1}{4T} \partial_\mu g_{\nu\lambda}(x) 
\psi^\mu \psi^\nu \dot x^{\lambda}
+ T m^2 \right\} \ . \quad
\eea
Clearly, there are no vertices with two or more gravitons in this picture. 
Using 
\bea
h_{\mu\nu}(x) \equiv g_{\mu\nu}(x) -\delta_{\mu\nu} =
\sum_{i=1}^{N} \epsilon_{\mu\nu}^{(i)} e^{ip_i \cdot x}
\eea
with the gravitational coupling constant included 
into the polarization tensors,
one gets the following general expression for the $N$-point effective
vertices
\bea
{1\over 2} \int_0^\infty {dT \over T} e^{-m^2 T}{2^{D\over 2}
\over (4 \pi T)^{D\over 2}}
\Big ( -{1\over 4T}\Big )^N \Bigl \la \prod_{i=1}^N 
\int_0^1 \!\! d \tau_i\,  V^{(i)}(\tau_i) \Bigr \ra
\eea
where the graviton vertex operator $V^{(i)} (\tau_i) $ is given by
\bea
V^{(i)}(\tau_i) = \epsilon_{\mu\nu}^{(i)} \Big ( \dot x^\mu \dot x^\nu +
a^\mu a^\nu +b^\mu c^\nu +\psi^\mu \dot \psi^\nu 
+\alpha^\mu \alpha^\nu
-i p_i^\lambda \psi_\lambda \dot x^\mu \psi^\nu \Big )(\tau_i)\, 
e^{ip_i \cdot x (\tau_i)}    \ .
\label{vo}
\eea

The explicit calculations of the one- and two-point vertex functions -- 
depicted in figures \ref{1ptf:m} and~\ref{2ptf:m}, respectively --  give 
the same results previously obtained from the coupling to the vielbein 
(after using relations~(\ref{onepoint}) and~(\ref{twopoint})). 
Note that external lines in figures \ref{1ptf:m} and~\ref{2ptf:m}
now refer to metric fluctuations.
Let us describe briefly these calculations.

\begin{figure}[ht]
\centering
\includegraphics{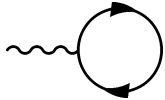}
\caption{\em Graviton tadpole.}
\label{1ptf:m}
\end{figure}

In the one-point function the connection term (i.e. the last term inside 
the round brackets of the vertex operator (\ref{vo}))
does not contribute, and the remaining terms lead to the same 
worldline integral obtained  previously
\bea
\int_0^1\! \! d\tau \, (\ddeld+\Delta_{gh}- \deld_{AF}-\Delta_{Agh})(0)  
 = \int_0^1\!\! d\tau \, (\ddeld+\Delta_{gh})(0)=1   \ .
\label{bfone}
\eea
In fact,  the propagator of the extra ghost field 
\bea
\la \a^\mu (\tau) \a^\nu(\sigma)\ra \!\! &=& \!\! 2T\, \delta^{\mu\nu}\,
   \Delta_{Agh}(\tau-\sigma)
\eea
where
\bea 
\Delta_{Agh}(\tau-\sigma)= \ddel_{AF}(\tau-\sigma)
=\delta_{A}(\tau-\sigma)
\label{ffgh}
\eea
cancels  with  $\deld_{AF}(\tau-\sigma) =-\ddel_{AF}(\tau-\sigma)$
due to the fermionic propagator.
This exemplifies the effect of the new ghost $\alpha^\mu$
which cancels a contraction arising from the $\psi^\mu \dot \psi^\nu $
term.
The final answer is
\bea
\Gamma^{\mu\nu}_{(0)} = 2^{{D \over 2}} {\delta^{\mu\nu} \over 4} 
{(m^2)^{D\over 2}\over (4 \pi)^{D\over 2} }
\Gamma\Big(-{D\over 2}\Big) \ .                               
\label{1ptfg}
\eea
As one might have expected, this result is $-2^{D\over 2}$ times 
the contribution of a scalar field \cite{Bastianelli:2002fv}: 
the minus sign is the usual one 
due to a fermionic loop, while $2^{D\over 2}$ is the number 
of degrees of freedom of a Dirac fermion in even $D$ dimensions. 

Let us now look at the two-point function. 
It corresponds to the single diagram of fig.~\ref{2ptf:m}, as 
in this scheme all vertices are linear in  the graviton field.
Again one may note that the $\alpha^\mu$ ghosts cancel all Wick contractions
arising from the $\psi \dot \psi $ term of the vertex operators
(notice that $\Delta_{AF}\deld_{AF} = -\Delta_{AF}\:\delta_{A}=0$).
Thus, only two non-vanishing contributions survive:
one from the square of the kinetic term of the bosonic sector
(i.e. $\sim (\dot x^2 + a^2 +bc)^2$);
 the other, transverse, from the square of the connection term.
\begin{figure}[ht]
\centering
\includegraphics{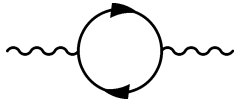}
\caption{\em Graviton self-energy.}
\label{2ptf:m}
\end{figure}
The final result is 
\bea
\Gamma_{(p,-p)} \!\! & = &\!\!  
{ 2^{{D \over 2}}\over 8 (4 \pi )^{D\over 2} } \left\{
\Gamma\Big(-{D\over 2}\Big)
  \biggl [ (m^2)^{{D\over 2}} ( R_1  -R_2 -S_1 -S_2)
  + (P^2)^{{D\over 2}}(S_1  +S_2)
  \biggr]  \right.                                   
                \nonumber \\
 &+&\!\! 
\left.  {1\over 4} \Gamma\Big(1-{D\over 2}\Big)\, p^2
 (P^2)^{{D\over 2}-1} S_2  \right\} \ .
\label{2ptfg}
\eea
This expression satisfies the expected gravitational Ward identities 
(for details see section A.5 of the appendix). 
The graviton self-energy  due to a massless fermion has been already
computed in  \cite{Capper:1973mv}, and agrees
with the massless limit of this general result 
\footnote{A similar result for a massive 
scalar field with minimal coupling to the scalar curvature 
can be found in \cite{Capper:1973bk} and agrees with the 
worldline result \cite{Bastianelli:2002fv}.
A calculation with standard Feynman rules for the scalar 
has been recently reported again in \cite{Gorbar:2002pw}.
It may be noticed how the worldline computation produces simpler 
and more compact expressions.}.

\section{Conclusions}

We have extended the worldline formalism to include fermionic 
fields coupled to gravity. To achieve this task we have found 
useful to 
study dimensional regularization on supersymmetric worldline
sigma models.
We have shown that dimensional regularization preserves
worldline supersymmetry in that no counterterms need to be 
added to the classical action to maintain supersymmetry.
This is in contrast to the time slicing regularization scheme, 
previously used for supersymmetric sigma models, which required
specific counterterms to restore supersymmetry.
Of course, final physical results are independent of the regularization 
scheme  adopted.
We have applied this set up to describe quantum  properties
of a Dirac fermion coupled to gravity.
Then, we have described the one-loop effective action
for a Dirac fermion coupled to gravity
in the worldline formalism, and computed the corresponding 
one- and two-point functions,
namely the one-loop fermionic contribution to the 
cosmological constant and graviton self-energy.
We have seen that one can use a formulation either in terms
of the vielbein or in terms of the metric, the latter being
much simpler as the coupling to gravity is linear
(and it avoids the introduction of the local Lorentz symmetry
related to a choice of the vielbein).
The computations are rather simple and demonstrate the efficiency 
of the worldline formalism in computing Feynman graphs
even in the presence of gravitational fields.
Our conclusion is that one can be confident and address more 
complicated processes using the worldline method.  In particular,
mixed photon-graviton amplitudes are under study~\cite{Bastianelli:2004zp}.

\vskip 1cm 

{\bf Acknowledgments}

We would like to thank Christian Schubert for useful
discussions and comments. We also thank the EC Commission 
for financial support via the FP5 Grant HPRN-CT-2002-00325.
We are indebted to Paolo Benincasa for pointing out an incorrect 
symmetrization of the two-point function in the vielbein basis, eq. 
(62) of the earlier version. 

\vfill \eject

\appendix

\section{Appendix}
\subsection{Covariant derivatives and curvature tensors}  
The covariant derivative for a vector with curved indices is 
$\nabla_\mu V^\nu= \partial_\mu V^\nu+ \Gamma_{\mu\lambda}^\nu V^\lambda$,
where $\Gamma_{\mu\lambda}^\nu= {1\over 2} g^{\nu\rho}
(\partial_\mu g_{\lambda\rho} +\partial_\lambda g_{\mu\rho}
-\partial_\rho g_{\mu\lambda})$
is the usual Christoffel connection.
The corresponding curvatures are defined by
\bea
[\nabla_\mu, \nabla_\nu] V^\lambda = 
R_{\mu\nu}{}^\lambda{}_\rho(\Gamma) V^\rho \ , \ \ \ 
R_{\mu\nu}= R_{\lambda\mu}{}^\lambda{}_\nu(\Gamma) 
\ , \ \ \ R= R^\mu{}_\mu > 0 \ {\rm on\ spheres.} 
\eea
The covariant derivative of a vector with flat indices is
$\nabla_{\!\!\mu} V^a= \partial_\mu V^a+ \omega_\mu{}^a{}_b V^b$,
where $\omega_\mu{}^a{}_b$ is the spin connection satisfying
the ``vielbein postulate'' $\nabla_{\!\! \mu}(\Gamma,\omega)\, e^a{}_\nu =0$.
The corresponding curvature is
\bea
[\nabla_\mu, \nabla_\nu] V^a = R_{\mu\nu}{}^a{}_b(\omega) V^b \ .
\eea
These curvatures are related by
\be
R_{\mu\nu}{}^\lambda{}_\rho(\Gamma)=
R_{\mu\nu}{}^a{}_b(\omega)e^\lambda{}_a e^b{}_\rho
\ .
\ee

\subsection{Propagators for bosons and related ghosts with vanishing 
Dirichlet boundary conditions} 

For quantum fields that vanish at $\tau=0,1$, we have the following
propagators
\bea
\la y^\mu (\tau) y^\nu(\sigma)\ra \!\! &=& \!\! 
- \beta \delta^{\mu\nu}
\Delta(\tau,\sigma)
\nonumber \\ 
\la a^\mu(\tau) a^\nu(\sigma)\ra \!\! &=& \!\! 
 \beta \delta^{\mu\nu}\Delta_{gh}
(\tau,\sigma) \label{propagdbc}
\nonumber \\ 
\la b^\mu(\tau) c^\nu (\sigma)\ra \!\! &=& \!\!
-2 \beta  \delta^{\mu\nu}\Delta_{gh}(\tau,\sigma)
\eea
with Green functions $\Delta$ and $\Delta_{gh}$ satisfying vanishing 
Dirichlet boundary conditions 
\bea
\Delta (\tau,\sigma)  \!\! &=& \!\! 
 \sum_{m=1}^{\infty}
\biggl [ - {2 \over {\pi^2 m^2}} \sin (\pi m \tau)
\sin (\pi m \sigma) \biggr ] =
(\tau-1)\sigma\, \theta(\tau-\sigma)+(\sigma-1)\tau\, \theta(\sigma-\tau)
\nonumber \\
\Delta_{gh}(\tau,\sigma) \!\! &=& \!\! 
\sum_{m=1}^{\infty}
2 \, \sin (\pi m \tau) \sin (\pi m \sigma) = \dddel(\tau,\sigma) =
\delta(\tau,\sigma)
\label{dbc2}
\eea
where $\theta(\tau-\sigma)$ is the standard step function and 
$\delta(\tau,\sigma)$ is the Dirac's delta function which vanishes at the 
boundaries $\tau,\sigma=0,1$.
These functions are not translationally invariant.

Their extensions to $d+1$ dimensions read 
\bea \hskip -.3cm
\Delta(t,s) \!\! &=& \!\! 
\int {d^d{\bf k}\over (2\pi)^d} \sum_{m=1}^\infty 
{-2\over (\pi m)^2+{\bf k}^2}\,
{\rm sin}(\pi m\tau)\, {\rm sin}(\pi m\sigma)\,
{\rm e}^{i{\bf k}\cdot ({\bf t}-{\bf s})}  
\label{4.1.18}
\\
\hskip -.3cm
\Delta_{gh}(t,s)
\!\! &=&\!\! \int {d^d{\bf k}\over (2\pi)^d} \sum_{m=1}^\infty 
2\, {\rm sin}(\pi m\tau)\, {\rm sin}(\pi m\sigma)\,
{\rm e}^{i{\bf k}\cdot ({\bf t}-{\bf s})} \nonumber
\\ &=& \!\!
 \delta (\t, \sigma)\, \delta^d ({\bf t} -{\bf s}) 
=\delta^{d+1}(t,s) \ .
\label{4.1.19}
\eea
Note that the function $\Delta(t,s)$ satisfies the relation
(Green's  equation)
\bea
\partial^\alpha\partial_\alpha \Delta(t,s) = \Delta_{gh}(t,s)=
\delta^{d+1}(t,s) \ .
\label{4.1.21}
\eea
The $d \to  0$ limits of these propagators 
reproduce the unregulated expressions.

\subsection{The ``string inspired'' propagators} 

The propagators we used in the worldline formalism are the 
``string inspired'' ones.
More specifically, on the circle the free kinetic term
for $x^\mu$ is proportional to $\p^2_{\tau}$  
and has a zero mode. Thus one splits
\bea
x^\mu(\tau) =x^\mu_0 +y^\mu(\tau), \qquad
x^\mu_0 =\int_0^1 d\tau\, x^\mu (\tau), \qquad
y^\mu(\tau) =\sum_{n\neq 0} y_n^\mu e^{2 \pi i  n \tau}
\eea
and the path integration measure becomes
\bea
 Dx= {1\over (4 \pi T)^{D \over 2}}\, d^Dx_0\,  Dy \ .
\label{mesplit}
\eea
The kinetic term for the quantum bosonic fields $y^\mu$ is invertible and
the corresponding free path integral is normalized to unity
\bea
\int Dy\ e^{-\int_0^1 d\tau {1\over 4 T} \dot y^2} =  1 \ .
\eea
The value of the free fermionic path integral defines implicitly its measure.
Using flat indices it reads
\bea
\int_{ABC} \!\! D\psi^a \; e^{-\int_{0}^{1} d\tau \; 
\frac{1}{2T} \psi_a\dot{\psi}^a}
= {\rm tr} (1)= 2^{D\over 2} \ .
\eea 
The propagators for the free fields are
\bea
\la y^\mu (\tau) y^\nu(\sigma)\ra \!\! &=& \!\!
- 2 T\, \delta^{\mu\nu}\, \Delta(\tau-\sigma)
\nonumber\\
\la a^\mu(\tau) a^\nu(\sigma)\ra \!\! &=& \!\!
  2T\, \delta^{\mu\nu}\, \Delta_{gh} (\tau- \sigma)
\nonumber\\
\la\, b^\mu(\tau)\, c^\nu(\sigma)\ra \!\! &=& \!\! -4T\, \delta^{\mu\nu}\,
\Delta_{gh}(\tau-\sigma)     \nonumber\\
\la \psi^a (\tau) \psi^b(\sigma)\ra \!\! &=& \!\!
 2 T\, \delta^{ab}\, \Delta_{AF}(\tau-\sigma)
\label{propag}
\eea
where $\Delta$, $\Delta_{gh}$ and $\Delta_{AF}$ are given by
\bea
&& \Delta (\tau-\sigma) = -\sum_{n\neq0}
 {1 \over {4 \pi^2 n^2}} e^{2 \pi i  n (\tau -\sigma )}
={1\over 2} |\tau-\sigma| -
{1\over 2} (\tau-\sigma)^2 -{1 \over 12}
\nonumber\\
&& \Delta_{gh}(\tau-\sigma) = \sum_{n=-\infty}^{\infty}
e^{2 \pi i n (\tau -\sigma)} = \delta_P(\tau-\sigma)
\label{ghost-prop} \\
&& \Delta_{AF} (\tau-\sigma) = \sum_{r\in Z+\frac{1}{2}}
 \frac{1}{2 \pi i r} e^{2 \pi i r(\tau -\sigma )}
=\frac{1}{2}\, \epsilon (\tau -\sigma)
\nonumber
\eea
and satisfy $\dddel=\Delta_{gh}-1=\delta_{P}-1$, 
$\ddel_{AF}=\delta_{A}$,
where $\delta_{P}$ and $\delta_{A}$ are the Dirac delta functions
on the space of periodic and antiperiodic functions on $[0,1]$,
respectively. 
All these free propagators are translationally invariant and have a well
defined parity under $(\tau -\sigma) \to (\sigma -\tau)$.

When using curved indices for the Majorana fermions $\psi^\mu$
there appears an extra set of bosonic ghosts $\alpha^\mu$.
Their propagators with ABC are
\bea
\la \psi^\mu (\tau) \psi^\nu(\sigma)\ra \!\! &=& \!\!
 2 T\, \delta^{\mu\nu}\, \Delta_{AF}(\tau-\sigma)
\nonumber\\
\la \alpha^\mu(\tau) \alpha^\nu(\sigma)\ra \!\! &=& \!\!
  2T\, \delta^{\mu\nu}\, \Delta_{Agh} (\tau- \sigma)
\label{propag-A}
\eea
where
\bea
\Delta_{Agh}(\tau-\sigma) = \sum_{r\in Z+{1\over2}}
e^{2 \pi i r (\tau -\sigma)} = \delta_A(\tau-\sigma) \ .
\eea
Clearly $\ddel_{AF} (\tau-\sigma) =\Delta_{Agh}(\tau-\sigma) 
=\delta_{A}(\tau-\sigma) $.

\subsection{Recursive formula for some worldline integrals} 

In the calculation of 1PI correlation functions via the worldline formalism
described in section~\ref{ss:world-line} one needs integrals of
the form
\bea
A_n = \int_0^1 \!\! d\tau \left(\tau-{1\over 2}\right)^{\!\!n} 
e^{-Tp^2(\tau-\tau^2)}
=\int_{-{1\over 2}}^{1\over 2} \! dx\, x^n e^{Tp^2(x^2 -{1\over 4})}~.
\eea
It is not difficult to prove the following recursive relations 
\bea
A_{2n+1} \!\! &=& \!\!
0\nonumber\\
A_{2n} \!\! &=& \!\! {1\over 2Tp^2}\left[ \left
({1\over 2}\right )^{\!\!2(n-1)}
-(2n-1) A_{ 2(n-1)}\right] 
\eea
and express all integrals in terms of $A_0$.

Recalling the definition of the gamma function
\bea
\Gamma(x) = \int_0^\infty dT\, T^{x-1} e^{-T}  
\eea
one can obtain the following result for the proper time integration
of $A_0$ 
\bea
\int_0^\infty dT\, T^{-x-1} e^{-m^2 T} A_0 = \Gamma(-x) (P^2)^x
\eea
where we have defined
\bea
(P^2)^x\equiv \int_0^1 d\tau\, (m^2+p^2(\tau-\tau^2))^x
 = (m^2)^x \: {}_2\!\!\: F_1 \left(-x,1;{3 \over 2};-{p^2 \over 4m^2} \right)~.
\eea
It is useful for the comparison with \cite{Capper:1973mv} to note that
\bea
\lim_{m^2 \to 0} (P^2)^x = (p^2)^x \: B(x+1,x+1)
\eea
Here we have used the hypergeometric function ${}_2\!\!\: F_1$ and the
Euler beta function $B$ .

\subsection{Ward identities} 
\label{section:wi}

A test for our results on one- and two-point functions
is provided by the Ward identities
due to general coordinate and local Lorentz invariances.
Local Lorentz symmetry $\delta e^a{}_\nu= \Lambda^a{}_b(x) e^b{}_\nu$
with arbitrary antisymmetric local $\Lambda_a{}_b(x) $
implies that 
\bea
{\delta \bar\Gamma[e] \over \delta e^a{}_\nu(x)} \Lambda^a{}_b(x)
e^b{}_\nu(x) =0 
\eea
which shows that the induced stress tensor 
$T_{\mu\nu}= {1\over e}
{\delta \bar\Gamma[e] \over \delta e_a{}^\mu} e_a{}_\nu$
is symmetric. In particular, it leads to the following identities for the one- 
and two-point functions in momentum space 
\bea
&&\bar\Gamma^{[a\nu]}_{(0)} =0\\
&&\bar\Gamma^{[a\mu] b \nu}_{(p,-p)}+\delta^{b[\mu}
\bar\Gamma^{a]\nu}_{(0)}=0 
\eea
where the square brackets denote weighted antisymmetrization. These identities
are satisfied by equations~(\ref{1ptf}) and~(\ref{2ptf}). Note in particular that 
the second identity allows for an antisymmetric contact term 
$\Delta_1\bar\Gamma^{[a\mu] b \nu}_{(p,-p)}$.

General coordinate invariance leads instead
to the conservation law for the induced energy-momentum tensor
\bea
\nabla_\nu^{(x)}\left( {1\over e(x)}
{\delta \bar\Gamma[e] \over \delta e^a_{\phantom{a}\nu}(x)}\
e^a_{\phantom{a}\mu}(x) 
\right) =0 \ .
\label{wwi}
\eea
Taking functional derivatives of this last expression produces
Ward identities that must be satisfied by the
one- and two-point functions 
\bea
&& p_\nu \tilde{\bar{\Gamma}}^{\mu\nu}_{(p)} =0    \\
&& p_\nu {\bar{\Gamma}}^{\mu\nu,\alpha\beta}_{(p,-p)}
+ p_\lambda  {\bar{\Gamma}}^{\alpha\lambda}_{(0)}\delta^{\mu\beta}
- p^\mu {\bar{\Gamma}}^{\alpha\beta}_{(0)} =0 \ .
\eea
It is easy to check that~(\ref{1ptf}) and~(\ref{2ptf}) do indeed satisfy the 
latter, while the former is rather straightforward ($p^\mu=0$
due to momentum conservation).

Alternatively, one can derive 
equivalent Ward identities for the effective action $\Gamma$ 
expressed as a functional of the metric 
and obtain
(equivalently, one may use relations (\ref{onepoint}-\ref{twopoint})) 
\bea
&& p_\mu \tilde{\Gamma}^{\mu\nu}_{(p)} =0    \\
&& p_\mu \Gamma^{\mu\nu,\alpha\beta}_{(p,-p)}
     +{1\over 2} p_\mu (\delta^{\nu\beta} \Gamma^{\mu\alpha}_{(0)}
     + \delta^{\nu\alpha} \Gamma^{\mu\beta}_{(0)} )
     -{1\over 2} p^\nu \Gamma^{\alpha\beta}_{(0)} =0~.
\eea
Also in this case it is simple to verify that 
eqs. (\ref{1ptfg}) and (\ref{2ptfg}) satisfy
these Ward identities.

\vfill \eject


\end{document}